\documentclass[preprint,12pt]{elsarticle}

\input epsf
\usepackage{amssymb}
\newcounter{bla}

\def\eqref#1{(\ref{#1})}

\def\erfc{{\rm erfc}}
\def\inverfc{{\rm inverfc}}

\def\erf{{\rm erf}}

\def\tfrac#1#2{{{\lower.6ex
\hbox{$\scriptstyle#1$}}\over
{\raise.7ex
\hbox{$\scriptstyle#2$}}}}

\def\erfc{{\rm erfc}}
\def\erf{{\rm erf}}

\def\bigO{{\cal O}}
\def\RR{{\mathbb R}}

\def\dsp#1{\displaystyle#1}

\def\Frac#1#2{\frac{\displaystyle{#1}}{\displaystyle{#2}}}

\def\bigO{{\cal O}}

\journal{Computer Physics Communications}

\begin{document}

\begin{frontmatter}



\title{{\bf GammaCHI}: a package for the inversion and computation of 
the gamma and chi-square cumulative distribution functions (central and noncentral)}


\author[1]{Amparo Gil}
\author[2]{Javier Segura}
\author[3]{Nico M. Temme}
\address[1]{Depto. de Matem\'atica Aplicada y Ciencias de la Comput. Universidad de Cantabria. 39005-Santander, Spain. e-mail: amparo.gil@unican.es}
\address[2]{Depto. de Matem\'aticas, Estad\'{\i}stica y Comput. Universidad de Cantabria. 39005-Santander, Spain}
\address[3]{ IAA, 1391 VD 18, Abcoude, The Netherlands\footnote{Former address: CWI, 1098 XG Amsterdam, The Netherlands} }

\begin{abstract}
A Fortran 90 module {\bf GammaCHI} for computing and inverting the gamma and chi-square cumulative distribution functions
(central and noncentral) is presented. The main novelty of this package are the
reliable and accurate inversion routines for the noncentral cumulative distribution functions.
Additionally, the package also provides routines for
computing the gamma function, the error function and other functions related to the gamma function. 
The module includes the routines {\bf cdfgamC}, {\bf invcdfgamC}, {\bf cdfgamNC}, {\bf invcdfgamNC}, 
 {\bf errorfunction}, {\bf inverfc}, {\bf gamma}, {\bf loggam}, {\bf gamstar} and {\bf quotgamm} 
 for the computation of the central gamma distribution function (and its complementary function), 
the inversion of the central gamma 
distribution function,  the computation of the noncentral gamma distribution function (and its complementary function), 
the inversion of the noncentral gamma 
distribution function,  the computation of the error function and its complementary function, the
inversion of the complementary error function, the computation of: the gamma function, 
 the logarithm of the gamma function,
the regulated gamma function and
 the ratio of two gamma functions, respectively.
\end{abstract}
\end{frontmatter}

{\bf PROGRAM SUMMARY}
  %Delete as appropriate.

\begin{small}
\noindent
{\em Manuscript Title:}                                       \\
GammaCHI: a package for the inversion and computation of 
the gamma and chi-square cumulative distribution functions (central and noncentral)\\
{\em Authors:}                                                \\
Amparo Gil, Javier Segura, Nico M. Temme                      \\
{\em Program Title:}                                          \\
Module {\bf GammaCHI}         \\
{\em Journal Reference:}                                      \\
  %Leave blank, supplied by Elsevier.
{\em Catalogue identifier:}                                   \\
  %Leave blank, supplied by Elsevier.
{\em Licensing provisions:}                                   \\
  %enter "none" if CPC non-profit use license is sufficient.
{\em Programming language:}                                   \\
  Fortran 90  \\
{\em Computer:}                                               \\
 Any supporting a FORTRAN compiler. \\
{\em Operating system:}                                       \\
 Any supporting a FORTRAN compiler.\\
{\em RAM:}                                              \\
   a few MB \\
{\em Number of processors used:}                              \\
  %If more than one processor.
{\em Keywords:}  \\
  % Please give some freely chosen keywords that we can use in a
  % cumulative keyword index.
  Gamma cumulative distribution function; chi-square cumulative distribution function;
inversion of cumulative distribution functions; 
error function;
complementary error function;
gamma function; logarithm of the gamma function; regulated gamma function; quotient of gamma functions. \\ 
{\em Classification:}                                         \\
  4.7 Other functions. \\
{\em External routines/libraries:}                            \\
  None. \\
{\em Subprograms used:}                                       \\
  %Fill in if necessary, otherwise leave out.
{\em Nature of problem:}\\
  %Describe the nature of the problem here.
The computation and inversion of gamma and chi-square cumulative distribution functions (central and noncentral)
as well as the computation of the error and gamma functions is needed in many problems of applied and mathematical physics. 
   \\
{\em Solution method:}\\
  The algorithms use different methods of computation depending
on the range of parameters: asymptotic expansions, quadrature methods, etc.
   \\
{\em Restrictions:}\\
  
In the inversion of the central
gamma/chi-square distribution functions, 
very small input function values  $P_{\mu}(x,y)$, $Q_{\mu}(x,y)$ 
(lower than $10^{-150}$) are not admissible. 

The admissible input parameter ranges for computing the noncentral cumulative
gamma distribution functions $P_{\mu}(x,y)$, $Q_{\mu}(x,y)$  in standard IEEE double
precision arithmetic are $0 \le x \le 10000$, $0 \le y \le 10000$, $0.5 \le \mu \le 10000$
and the related parameter ranges for the noncentral chi-square cumulative
distribution function. In the inversion of the noncentral
gamma/chi-square distribution functions, very small input function values  $P_{\mu}(x,y)$, $Q_{\mu}(x,y)$ 
(lower than $10^{-25}$ and $10^{-35}$, respectively) are not admissible. 
   \\
{\em Running time:}\\
  %Give an indication of the typical running time here.
It varies depending on the function and the parameter range.
   \\
\end{small}

\section{Introduction}

The Fortran 90 module {\bf GammaCHI} provides reliable and fast routines
for the inversion and computation of the gamma and chi-square distribution functions. 
These functions  appear in many problems of applied probability including, of course, a large number
of problems in Physics. 
 The module also includes routines for the computation of the
gamma function, the error function and its complementary function, the inverse of the
complementary error function,  the logarithm of the gamma function,
the regulated gamma function and
 the ratio of two gamma functions.

The main novelty of the package {\bf GammaCHI} are the inversion routines for the noncentral
cumulative distribution functions. These algorithms are based on the 
asymptotic methods presented in \cite{Gil:2013:IMQ} although improvements 
in the estimation of the initial values for small values of the parameter $\mu$
are included in the present algorithms. Among other applications, the computation and the inversion 
of the noncentral cumulative distribution functions included in {\bf GammaCHI} appear in the analysis of signal detection 
in different physical scenarios such as  optics \cite{tang:1993:cbe}, radiometry \cite{mills:1996:cvr} or  
 quantum detection \cite{helstrom:1976:qde}, \cite{vasylyev:2012:prl}.
Also, the inversion routines can be used to generate normal 
variates as well as central and noncentral gamma (or chi-square) 
variates, which
can be useful, for example, in Monte Carlo simulations.

\section{Theoretical background}

\subsection{Central gamma distribution function}

The incomplete gamma functions are defined by

\begin{equation}\label{eq:int01}
\gamma(a,x)=\int_0^x t^{a-1} e^{-t}\,dt, \quad
\Gamma(a,x)=\int_x^{\infty} t^{a-1} e^{-t}\,dt.
\end{equation}

Let us define

\begin{equation}\label{eq:int02}
P(a,x)=\frac{1}{\Gamma(a)}\gamma(a,x), \quad
Q(a,x)=\frac{1}{\Gamma(a)}\Gamma(a,x),
\end{equation}
where we assume that $a$ and $x$ are real positive numbers.

The functions $P(a,x),\, Q(a,x)$  are the central gamma distribution function and its complementary function, respectively.
These distribution functions which appear in many problems of applied probability
include, as particular cases, the standard chi-square probability functions $P(\chi^2| \nu)$ and
$Q(\chi^2 | \nu)$ with parameters $a=\nu/2$ and $x=\chi^2/2$.  

The central gamma distribution functions satisfy the complementary relation

\begin{equation}\label{eq:cr}
P(a,x)+Q(a,x)=1.
\end{equation}

   In hypothesis testing, and using the chi-square distribution, it is usual to consider 
the problem as $Q(\chi^2|\nu)=\alpha$; $\alpha$ is the probability that a variable distributed according to the chi-square
distribution with $\nu$ degrees of freedom exceeds the value $\chi^2$.
Two kinds of problems are usually considered: a) computing the confidence 
level $\alpha$ given an experimentally determined chi-square value $\chi^2$;
 b) computing percentage points for
certain values of $\alpha$ and for different degrees of freedom $\nu$ (the standard tables which are common in
statistics text-books usually include this information). 
The first problem involves the direct computation of the cumulative distribution function while
the second one is an inversion problem.  

 The module {\bf GammaCHI} includes routines both for the direct computation
of the distribution functions (central gamma/chi-square) and their inversion.
Regarding the direct computation, the routine first computes $\min\{P(a,x), Q(a,x)\}$, and the other one by using \eqref{eq:cr}. Computing the $Q(a,x)$ function simply 
as $1-P(a,x)$ when $P(a,x)$ is close to 1 can lead to serious cancellation problems, which are avoided in our algorithms.

  On the other hand, the inversion routine solves
 the equations
\begin{equation}\label{eq:int05}
P(a,x)=p,\quad Q(a,x)=q, \quad 0<p,\ q<1,
\end{equation}
for a given value of $a$.

 The algorithm for computing the gamma distribution function and
its inversion is described in \cite{Gil:2012:IGR}. 
The computation of the distribution functions is based on the use of Taylor expansions,
continued fractions or uniform asymptotic expansions, depending on the parameter
values. For the inversion, asymptotic expansions in combination with high-order Newton methods
are used.

The version included
in the package {\bf GammaCHI} for the inversion includes some improvements in its performance for small values
of $p$ and $q$. For such small values, some of the elementary bounds for incomplete gamma functions given
in \cite{Segura:2014:MPA} provide accurate enough approximations for estimating sufficiently accurate starting values
for the Newton iterative process.

\subsection{Noncentral gamma distribution function}

The noncentral gamma cumulative distribution functions can be defined as

\begin{equation}\label{eq:def02}
\begin{array}{l}
\dsp{P_{\mu} (x,y)=e^{-x}\sum_{k=0}^{\infty} \frac{x^k}{k!}  P(\mu +k,y),}\\[8pt]
\dsp{Q_{\mu} (x,y)=e^{-x}\sum_{k=0}^{\infty} \frac{x^k}{k!}  Q(\mu +k,y),}
\end{array}
\end{equation}
where $ P(\mu,y)$  and $Q(\mu,y)$ are the central gamma cumulative distribution function and its complementary function,
respectively; $x$ is called the noncentrality parameter of the distribution functions.

As in the case of the central distribution functions, the noncentral functions satisfy the relation

\begin{equation}
P_{\mu} (x,y)+Q_{\mu} (x,y)=1.
\end{equation}

On the other hand, if $\chi^2_n(\lambda)$ is a random
variable with a noncentral chi-square distribution with $n>0$ degrees of freedom and noncentrality parameter
$\lambda \ge 0$, then the lower and upper tail probabilities
for the distribution function of $\chi^2_n(\lambda)$ are given by

\begin{equation}\label{eq:chi1}
\begin{array}{l}
\mbox{Prob}(\chi^2_n(\lambda)<t) =e^{-\lambda/2}\dsp{\sum_{k=0}^{\infty} } \Frac{(\lambda/2)^k}{k!}P(n/2+k,\,t/2)  ,\\[8pt]
\mbox{Prob}(\chi^2_n(\lambda)>t) =e^{-\lambda/2}\dsp{\sum_{k=0}^{\infty} } \Frac{(\lambda/2)^k}{k!} Q(n/2+k,\,t/2),
\end{array}
\end{equation}
from which, and comparing with (\ref{eq:def02}), it follows the simple relation between distribution functions: 
if the random variable $Y$ has a noncentral gamma distribution with parameters $a$ and noncentrality parameter 
$\lambda$, then $X=2Y$ has a noncentral chi-square distribution with parameter $n=2a$ and with noncentrality
parameter $2\lambda$.    

Algorithms for computing the functions $P_{\mu}(x,y)$ and $Q_{\mu} (x,y)$ for a
 large range of the parameters $\mu$ ($\mu\ge 1$), $x$, $y$ are described in \cite{Gil:2012:CMQ}. 
The methods include
series expansions in terms of the incomplete gamma functions,
 recurrence relations, continued fractions, asymptotic expansions, and numerical quadrature. 
In the module {\bf GammaCHI} the lower $\mu$ range of computation of the algorithm
for computing the noncentral gamma cumulative distribution function  is extended to include the interval
$\frac{1}{2}\le \mu<1$. For extending the algorithms to this interval, we use a single step of the following
three-term homogeneous 
recurrence relation (Eq.14 of \cite{Gil:2012:CMQ})

\begin{equation}
\label{TTRR}
y_{\mu+1} -(1+c_\mu) y_\mu + c_\mu y_{\mu-1}=0,\quad c_\mu=\sqrt{\frac{y}{x}}\Frac{I_\mu \left(2\sqrt{xy}\right)}{I_{\mu-1}\left(2\sqrt{xy}\right)}.
\end{equation}

Both $Q_\mu (x ,y)$ and $P_\mu (x ,y)$ satisfy (\ref{TTRR}). 
In this expression, ratios of Bessel functions appear. These ratios are efficiently computed using 
a continued fraction representation.

Regarding the inversion of the noncentral gamma/chi-square distribution
functions, two different kinds of inversion problems can be considered: 

\begin{description}
\item[Problem~1:]
Find $x$ from the  equation 
\begin{equation}\label{eq:inv01}
Q_\mu(x,y)=q,
\end{equation}
with fixed $y$, $\mu$ and $q$.

\item[Problem~2:]
Find $y$ from the  equation 
\begin{equation}\label{eq:inv02}
P_\mu(x,y)=p,
\end{equation}
with fixed $x$, $\mu$ and $p$.
\end{description}

The inversion of  $Q_\mu(x,y)$ with respect to $x$ corresponds to 
the problem of inverting the distribution function with
respect to the noncentrality parameter given the upper tail probability.
On the other hand, the inversion of $P_{\mu}(x,y)$ with respect to $y$ with fixed $x$
corresponds to the problem of computing the $p$-quantiles of the distribution function.
For noncentral chi-square distributions, the inversion with respect to $y$ allows the computation of a variate with a given 
number of degrees of freedom and a given noncentrality parameter.

In the inversion with respect to $x$, it is important to note that not for all
the possible values of the pair $(y_0,q_1)$ there is a solution of the equation

\begin{equation}
Q_\mu(x,y_0)=q_1.
\end{equation}

In particular, the values of $q_1$ such that $q_1<q_0$, where $q_0=Q_\mu(0,y_0)=Q_\mu(y_0)$, do not make sense
because $Q_{\mu}(x,y)$ is increasing as a function of $x$.

Both in the inversion with respect to $x$ and $y$, a combination of asymptotic expansions and secant methods can be used, as discussed
in \cite{Gil:2013:IMQ}.  A Fortran 90 implementation is now available in the module {\bf GammaCHI}.
As for the direct computation of function distribution values, in this implementation
the range of computation for the inversion of the
gamma cumulative distribution function is extended to include the interval $\frac{1}{2}\le \mu<1$. The computation
in this interval is made by using the double asymptotic property of the expansions derived in \cite{Gil:2013:IMQ}:
the formulas obtained in \cite{Gil:2013:IMQ} can all be used for small values of $\mu$ if $\xi=2\sqrt{xy}$ is large. 
More details can be found in the appendix. For small $q$ this is suplemented with estimations for the case $\mu=1/2$,
which can be written in terms of error functions; we use the result of the inversion for the case $\mu=1/2$ to
estimate a starting value when $\mu$ is close to $1/2$. It is easy to check that 
$$Q_{1/2}(x,y)=\Frac{1}{2}\left(\erfc (\sqrt{x}+\sqrt{y})+\erfc (\sqrt{y}-\sqrt{x}) \right)$$ and for this function, computing
derivatives is simple and one can apply the fourth order method described in \cite{Segura:2015:TSM}.   
 
\subsection{Error function and its complementary function} \label{introerror}

The error function $\erf(x$) \cite{temme:2010:error} is defined by the integral

\begin{equation}
\label{eq:errorf01}
\erf (x)= \Frac{2}{\sqrt{\pi}} \displaystyle\int_0^x e^{-t^2}\,dt.
\end{equation} 

Its complement with respect to 1 is denoted by
  
\begin{equation}
\label{eq:errorf02}
\erfc (x)= 1- \erf (x)=\Frac{2}{\sqrt{\pi}} \displaystyle\int_x^\infty e^{-t^2}\,dt.
\end{equation}

The complementary error function is very closely related to the normal distribution functions,
which are defined by

\begin{equation}
\label{eq:errorf03}
P(x)=\Frac{1}{\sqrt{2\pi}} \displaystyle\int_{-\infty}^x e^{-t^2/2}\,dt,\,\,
Q(x)=\Frac{1}{\sqrt{2\pi}} \displaystyle\int_x^{\infty} e^{-t^2/2}\,dt
\end{equation}
with the property $P(x)+Q(x)=1$.

From the definition of the complementary error function it follows that

\begin{equation}
\label{normal}
P(x)=\Frac{1}{2}\erfc (-x/\sqrt{2}),\,\,Q(x)=\Frac{1}{2}\erfc (x/\sqrt{2}).
\end{equation}

The algorithm for the error functions is based on the use of rational approximations \cite{cody:1969:RAT} and
it includes the possibility of computing the scaled complementary error function 
$e^{x^2} \erfc (x)$. 
This is important to avoid underflow problems in numerical
computations of the complementary error function
(the dominant exponential behavior of the function is $e^{-x^2}$). In general,
 it is quite useful (when possible) to define scaled values of mathematical functions 
with the dominant exponential behavior factored out \cite[\S12.1.3]{Gil:2007:NSF}.  

Our module includes also the computation of the inverse of the
complementary error function $y=\inverfc\,x$. Using the relation (\ref{normal}), this routine 
can be also useful in the generation of normal variates.

For values of $x$ close to $1$, the following expansion is
used 

\begin{equation}\label{eq:inverfc1}
\inverfc\,x=t+\tfrac13t^3+\tfrac{7}{30}t^5+\tfrac{127}{630}t^7+\ldots, \quad 0<x<2.
\end{equation}
with $t=\frac12\sqrt\pi\,(1-x)$. For these and more coefficients, see \cite{Strecok:1968:OCI}.

\subsection{Gamma function}

Three standard definitions of the gamma function are usually considered:

\begin{equation}
\label{def1}
\Gamma (z) =\displaystyle\int_0^{\infty} e^{-t} t^{z-1}\,dt,\,\,\mbox{Re }z >0\,\,\,\mbox{(Euler)}.
\end{equation}

\begin{equation}
\label{def2}
\Gamma (z) =\displaystyle\lim_{n \rightarrow \infty} 
\Frac{n! \,n^z}{z(z+1)\cdots(z+n)},\,\,z \neq 0,-1,-2,...\,.\,\,\,\mbox{(Euler)}.
\end{equation}

\begin{equation}
\label{def3}
\Frac{1}{\Gamma (z)} = z e^{\gamma z}\displaystyle\prod_{n=1}^{\infty} \left(1+\Frac{z}{n}\right) e^{-z/n}\,\,\,\mbox{(Weierstrass)},
\end{equation}
with $\gamma=0.57721...$, Euler's constant. The equivalence of these definitions is proved in 
\cite{Hochstadt:1971:FMP}.

Important relations are

\begin{equation}
\Gamma(z+1)=z \Gamma(z),\,\,\,\mbox{(recursion)},
\end{equation}
and

\begin{equation}
\label{reduc}
\Frac{1}{\Gamma (1+z)\Gamma (1-z)} = \Frac{\sin \pi z}{ \pi z}, 
\end{equation}
which is called the reflection formula. This formula
is important for range reduction and it is easily proved by using (\ref{def3}).

For computing the gamma function, we use recursion for small values of $x$ and 
the relation given in (\ref{asymp}) for larger values.
The special cases $x=n$ and $x=n+\frac{1}{2}$ are treated separately.  

\subsubsection{Logarithm of the gamma function}

 The following representation of $\log \Gamma (z)$ is very
important for deriving expansions for large values of $|z|$:

\begin{equation}
\label{asymp}
\log \Gamma (z) = \left(z-\tfrac{1}{2}\right) \log z -z +\tfrac{1}{2} \log (2 \pi) + S(z).
\end{equation}

For large $|z|$, $S(z)$ gives a small correction with respect to the remaining
terms of the right-hand side. Neglecting $S(z)$ gives the well-known Stirling formula

\begin{equation}
\label{stir}
\Gamma (z) \sim z^z e^{-z} \left(\Frac{2 \pi}{z}  \right)^{1/2},\,\,\,z \rightarrow \infty.
\end{equation}

$S(z)$ can be written in terms of an expansion involving Bernoulli numbers: 

\begin{equation}
\label{szexp}
S(z)=\displaystyle\sum_{n=0}^{N-1} \Frac{B_{2n+2}}{(2n+1)(2n+2)z^{2n+1}}+E_N(z).
\end{equation} 

The Bernoulli numbers $B_n$ are given by $B_n=B_n(0)$. The first few are
$B_0=1,\,B_1=-\Frac{1}{2},\,B_2=\Frac{1}{6},\,B_3=0,\,B_4=-\Frac{1}{30}$. More information
on Bernoulli polynomials and Bernoulli numbers can be found in 
\cite[\S24.2(i)]{olver:2010:NHB}.

A bound for $|E_N|$ in \eqref{szexp} can be obtained as

\begin{equation}
\label{bound}
|E_N| \le \Frac{B_{2N+2}\, K(z)}{(2N+1)(2N+2)}|z|^{-2N-1},
\end{equation}
where 

$$
K(z) =\sup_{u \ge 0} \left|\Frac{z^2}{z^2+u^2}   \right|.
$$

If $|{\rm arg}\,z| < \pi/4$, then $K(z) =1$. For real positive $z$, $E_N(z)$ is
less in absolute value than the first term neglected in (\ref{szexp})
and it has the same sign. From (\ref{bound}) it follows that
$E_N(z) = {\cal{O}} (z^{-2N-1})$ for $\mbox{Re } z \rightarrow + \infty$. Inserting
the values of the first Bernoulli numbers we arrive at the representation
(Stirling's series):

\begin{equation}
\label{stirling}
\begin{array}{@{}r@{\,}c@{\,}l@{}}
\log \Gamma (z) &=& \left(z-\Frac{1}{2}\right) \log z -z + \Frac{1}{2} \log (2 \pi) \ +\\[8pt]
&& \Frac{1}{12z} - \Frac{1}{360z^3}+ \Frac{1}{1260z^5}-\Frac{1}{1680z^7}+ {\cal O} (z^{-9}).
\end{array}
\end{equation}

The algorithm for the logarithm of the gamma function makes use of the representation
(\ref{asymp}). The computation of $S(z)$ is based on the
use of the sum (\ref{asymp}), Chebyshev series and a minimax rational approximation.
Coefficients of rational approximations are taken from the the tables in \cite[\S6.6]{Hart:1968:CAP}. 
On some intervals we use Chebyshev expansions, with coefficients derived by using high precision Maple codes.

\subsection{The function $\Gamma^*(x)$}

This function (the regulated gamma function) is defined by 

\begin{equation}\label{eq:moc05}
\Gamma^*(x)=\frac{\Gamma(x)}{\sqrt{2\pi/x}\, x^xe^{-x}},\quad x>0.
\end{equation}

When $x$ is large, this function can be very important in algorithms where the function
$\Gamma (x)$ is involved  because
$\Gamma^*(x)=1 +{\cal{O}}(1/x)$,
as can be seen from its asymptotic expansion (Stirling series):

\begin{equation}
\Gamma^*(x)\sim 1+\tfrac{1}{12}x^{-1}+ \tfrac{1}{288}x^{-2}+\ldots, \quad x \to \infty.
\end{equation}

The algorithm for computing $\Gamma^*(x)$, uses the relation $\Gamma^*(x)=\exp(S(x))$ for $x>3$, where $S(x)$  
is given in (\ref{szexp}). The computation of $S(x)$ was described in the previous section.

\subsection{The quotient of two gamma functions}

 The quotient of two gamma functions appears frequently in applications. From a computational
point of view, problems can arise when trying to compute directly the ratio 
of functions in the case when the arguments of both functions are large. 
For that reason, a key point in our algorithm for computing
the quotient of two gamma functions is the use of the following asymptotic expansion \cite[p.~68]{Temme:1996:SFA}:

\begin{equation}
\label{ratgam}
\Frac{\Gamma(x+a)}{\Gamma(x+b)} \sim w^{a-b} \displaystyle\sum_{n=0}^{\infty} (-1)^n C_n 
\Frac{\Gamma (b-a+2n)}{\Gamma (b-a)}\Frac{1}{w^{2n}},\,\,\mbox{as } x \rightarrow \infty,
\end{equation}
with $w=x+(a+b-1)/2$ and $b>a$.

The first five coefficients $C_n$ appearing in (\ref{ratgam}) are:

\begin{equation}
\begin{array}{lcl}
 C_0&=&1,\\
 C_1&=&\Frac{\rho}{12},\\
 C_2&=&\Frac{\rho}{1440}+\Frac{\rho^2}{288},\\
 C_3&=&\Frac{\rho}{90720}+\Frac{\rho^2}{17280}+\Frac{\rho^3}{10368},\\
 C_4&=&\Frac{\rho}{4838400}+101\Frac{\rho^2}{87091200}+
\Frac{\rho^3}{414720}+\Frac{\rho^4}{497664},
\end{array}
\end{equation}
with $\rho=(a-b+1)/2$.

\section{Overview of the software structure}

The Fortran 90 package includes the main module {\bf GammaCHI}, which includes
as public routines the following functions and routines: {\bf cdfgamC}, {\bf invcdfgamC}, {\bf cdfgamNC}, {\bf invcdfgamNC}, 
 {\bf errorfunction}, {\bf inverfc}, {\bf gamma}, {\bf loggam}, {\bf gamstar} and {\bf quotgamm}. 

In the module {\bf GammaCHI}, the auxiliary 
module {\bf Someconstants} is used. This is a module for the 
computation of the main constants used in 
  the different routines.

\section{Description of the individual software components}

\begin{enumerate}

\item  {\bf cdfgamC}:

The calling sequence of this routine is
\begin{verbatim}
    cdfgamC(ichi,a,x,p,q,ierr)
  \end{verbatim}
  where the input data are: $ichi$ (flag),  $a$ and $x$ (arguments of the distribution function). 
$ichi$ is a flag for the choice of a gamma or chi-square cumulative distribution function:
when $ichi=1$, the arguments are of a gamma distribution function; when $ichi=2$, the arguments
correspond to a central chi-square distribution function.

The outputs of the function are error flag $ierr$, the function distribution value $p$ and its complementary function
$q$. The possible values of the error flag are: $ierr=0$, successful 
computation; 
$ierr =1$, computation failed due to overflow/underflow; 
$ierr=2$, arguments out of range.

\item  {\bf invcdfgamC}:

The calling sequence of this routine is
\begin{verbatim}
    invcdfgamC(ichi,a,p,q,xr,ierr)
  \end{verbatim}
  where the input data are: $ichi$ (flag), $a$, $p$ and $q$. $ichi$ is a flag for the choice 
of a gamma or chi-square cumulative distribution function:
when $ichi=1$, a gamma distribution function is inverted; when $ichi=2$, a central chi-square 
distribution function is inverted.

The outputs of the function are $xr$ (the solution of the equations
 $P(a,xr)=p$ and $Q(a,xr)=q$) and the error flag $ierr$. The possible values 
of the error flag are: $ierr=0$,  computation successful;
$ierr=1$,  overflow problems in one or more steps of the 
                computation;
             $ierr=2$,  the number of iterations in the Newton method
                       reached the upper limit $N=35$;
              $ierr=3$,  any of the arguments of the function is 
               out of range.  

\item  {\bf cdfgamNC}:

The calling sequence of this routine is
\begin{verbatim}
    cdfgamNC(ichi,mu,x,y,p,q,ierr)
  \end{verbatim}
  where the input data are: $ichi$ (flag), $mu$, $x$ and $y$ (arguments of the distribution function). 
The input $ichi$ input is a flag for the choice of a gamma or chi-square noncentral cumulative distribution function:
when $ichi=1$, the arguments are of a gamma distribution function; when $ichi=2$, the arguments
correspond to a noncentral chi-square distribution function.

The outputs of the function are: the function distribution value $p$ and its complementary function
$q$ and  an error flag $ierr$. The possible values of the error flag are: $ierr=0$, successful 
computation; 
$ierr =1$, computation failed due to overflow/underflow; 
$ierr=2$, arguments out of range. 

\item  {\bf invcdfgamNC}:

The calling sequence of this routine is
\begin{verbatim}
    invcdfgamNC(ichi,icho,mu,p,q,yx,xy,ierr)
  \end{verbatim}
where the input data are: $ichi$, $icho$, $mu$, $p$, $q$ and $yx$.
$ichi$ is a flag for the choice of a noncentral gamma or
 chi-square cumulative distribution function:
when $ichi=1$, the arguments are of a gamma distribution function; when $ichi=2$, the arguments
correspond to a central chi-square distribution function. The input $icho$ is a flag for the choice
of the inversion process: if $icho=1$, $x$ is computed in the
                          equations $Q_{mu}(x,yx)=q$, $P_{mu}(x,yx)=p$. 
              If $icho=2$, $y$ is computed in the
                           equations $Q_{mu}(yx,y)=q$, $P_{mu}(yx,y)=p$;
$q$ and $p$ are the values of the cumulative distribution functions;
$mu$ and $yx$ are arguments of the distribution functions. 

The outputs of the routine are the value obtained in the inversion $xy$ and an error flag
$ierr$. The possible values of the error flags are: $ierr=0$, computation successful;
              $ierr=1$, in the inversion process with respect to the 
                      non-centrality parameter ($x$-variable), the input 
                      values $(y,q)$ do not make sense (i.e. the 
                      inequality $q<Q_{mu}(0,y)$ holds); $ierr=2$, at least one of the gamma distribution 
                      function values needed in the inversion process
                      cannot be correctly computed;  $ierr=3$, the number of iterations in the secant method
                      reached the upper limit; $ierr=4$, any of the input arguments is 
                      out of range.  

\item  {\bf errorfunction}:

The calling sequence of this function is
  \begin{verbatim}
   errorfunction(x,erfc,expo)
  \end{verbatim}
  where the input data are: $x$ (argument of the function), erfc and expo
(logical variables).  

The logical variables erfc and expo mean the following:

\begin{description}
\item When erfc=.true. and expo=.false., the function computes the complementary error function erfc($x$). 
\item When erfc=.true., expo=.true. and $x>0$, the function computes the scaled complementary
error function $e^{x^2}$erfc($x$). 
\item When erfc=.false. and expo=.false., the function computes the error function erf($x$).
\end{description}

\item {\bf inverfc}:

The calling sequence of this function is
  \begin{verbatim}
   inverfc(x)
  \end{verbatim}
The input argument is $x$, a positive real number.
This function computes the inverse of the complementary error function.

\item  {\bf gamma}:

The calling sequence of this function is
\begin{verbatim}
   gamma(x)
  \end{verbatim}
This function computes the Euler gamma function $\Gamma(x)$, for $x$ real.

\item  {\bf loggam}:

The calling sequence of this function is
  \begin{verbatim}
   loggam(x)
  \end{verbatim}
This function computes $\log(\Gamma(x))$, for $x>0$.

\item {\bf gamstar}:
 
The calling sequence of this function is
  \begin{verbatim}
   gamstar(x)
  \end{verbatim}
This function computes for positve $x$ the regulated gamma function $\Gamma^*(x) = \Frac{\Gamma(x)}{\sqrt{2\pi/x}\, x^xe^{-x}}$.
 
\item  {\bf quotgamm}:

The calling sequence of this function is
  \begin{verbatim}
   quotgamm(x,y)
  \end{verbatim}
This function computes $\Frac{\Gamma(x)}{\Gamma(y)}$ for $x$, $y$ real values.

\end{enumerate}

\section{Testing the algorithms}

 For testing the algorithm for the computation of the central gamma distribution functions, 
we use the recurrence relations (see \cite[\S8.8]{Paris:2010:IGF})

\begin{equation}\label{eq:moc08}
P(a+1,x)=P(a,x)-D(a,x),\quad Q(a+1,x)=Q(a,x)+D(a,x),
\end{equation}
where
\begin{equation}\label{eq:moc03}
D(a,x)=\frac{x^a e^{-x}}{\Gamma(a+1)}.
\end{equation}

The recurrence relations (\ref{eq:moc08}) are tested for a large number of random points in
the input parameter domain $(a,x),\,x\le 0,\,a>10^{-300}$. The aimed accuracy is close to $10^{-13}$ in most cases.  
For small values of $a$, the accuracy of the values computed by our routine {\bf cdfgamC}
 is even higher, near $10^{-15}$ as can be seen in Table 1. In this table, a comparison of a few
values against those obtained using the Matlab function {\bf gammainc} included in the 
release 2013a, is shown. It should be noted that versions of {\bf gammainc} included in some previous
Matlab releases showed some loss of accuracy for these same values.

\begin{table}
\label{table1}
$$
\begin{array}{cccc}
a  & x  & Q(a,x)  &  gammainc(x,a,'upper')    \\
  \hline
 10^{-250} & 6.3\,10^{-15} & {\bf 3.212101109661167} \, 10^{-249} &   3.212101109661167\, 10^{-249}  \\
 10^{-250} & 7.1\,10^{-7} &  {\bf 1.35807859120093}93  \, 10^{-249} &  1.35807859120093 \, 10^{-249} \\
 10^{-250} & 0.01 &   {\bf 4.03792957653811}35  \, 10^{-250} &  4.037929576538114 \,  \, 10^{-250} \\ 
 10^{-14} & 6.3\,10^{-15} & {\bf 3.21210110966065}1 \, 10^{-13} &   3.212101109660652\, 10^{-13}  \\
 10^{-14} & 7.1\,10^{-7} &  {\bf 1.358078591200848}  \, 10^{-13} &  1.358078591200848  \, 10^{-13} \\
 10^{-14} & 0.01 &   {\bf 4.0379295765380}405  \, 10^{-14} &  4.037929576538039 \,  \, 10^{-14} \\ 
\hline
\end{array}
$$
{\footnotesize {\bf Table 1}. Test of the routine {\bf cdfgamC} for small values
of $a$ and $x$: comparison of the values obtained for $Q(a,x)$ using our routine
{\bf cdfgamC} and the Matlab function {\bf gammainc} included in the 2013a release.
It should be noted that our value for $x=0.01$, $a=10^{-14}$ agrees in 16 digits with
the value obtained with the Mathematica function {\bf GammaRegularized[a,x]}:
 $4.03792957653804043... \, 10^{-14 }$.}
\end{table}

For the inversion algorithm, testing is made by checking that the composition of the functions with 
their inverse is the identity.
  Some tests for the inversion routine  {\bf invcdfgamC} are shown
in Table 2. This table shows the relative error for the inversion
problem of the central gamma distribution function given in Eq.~(\ref{eq:int05}) for different
values of $p$ and $a$.

 \begin{table}
\label{table2}
$$
\begin{array}{cccccc}
  &   & a &  & \\
  \hline
p & 0.05 & 1 & 10 & 100 & 1000 \\
\hline
0.0001 &  4\,10^{-16}& 4.1\,10^{-16} & 6.5\,10^{-15}  & 8.1\,10^{-16} & 5.3\,10^{-15} \\
0.01 & 2\,10^{-16} & 5.2\,10^{-16}  & 2.9\,10^{-15} & 1.2\,10^{-15}   & 1.9\,10^{-15}  \\
0.1 & 2.8\,10^{-16} & 1.4\,10^{-16}  & 0  & 1.5\,10^{-15}  & 8.3\,10^{-16}\\
0.3 & 0 & 1.8\,10^{-16} & 3.7\,10^{-16} & 7.4\,10^{-16} & 7.4\,10^{-16} \\
0.5 & 10^{-16} & 0  & 0  & 1.1\,10^{-16} & 2.2\,10^{-16} \\
0.7 & 0 & 1.6\,10^{-16}  & 0  & 1.6\,10^{-16}  &   7.9\,10^{-16}  \\
0.9 & 0 & 0 & 0  & 0  &  3.7\,10^{-16}  \\
0.9999 & 0 & 0 & 0  & 0  &  0  \\
\end{array}
$$
{\footnotesize {\bf Table 2}. Relative errors $|P(a,x_r)-p|/p$ for several
values of $p$ and $a$.  }
\end{table}

Further tests for the central gamma distribution functions can be found in \cite{Gil:2012:IGR}.
 
Testing of the noncentral gamma distributions can be made by checking the deviations
from $1$ of the relation

\begin{equation}
\Frac{(x-\mu)Q_{\mu+1} (x,y)+ (y+\mu)Q_{\mu} (x,y)}{xQ_{\mu+2}(x,y)+yQ_{\mu-1} (x,y)}=1.      
\label{errRR}
\end{equation}

Equation (\ref{errRR}) is used when $y \ge  x+\mu$ and the same expression but for $P_{\mu}(x,y)$ 
when $y < x +\mu$.
The aimed accuracy for the computation of the noncentral gamma
cumulative distribution functions is close to $10^{-11}$ in most cases  in the 
parameter domain $(x,\,y,\,\mu)$ with $0 \le x \le 10000$, $0 \le y \le 10000$ and $0.5 \le \mu \le 10000$. 

We have compared our routines for the noncentral distributions against the package {\bf ncg} in R. The description
of the algorithms included in this package is given in \cite{Oliveira:2013:CNC}.
Algorithms (but not associated software) for the computation of the noncentral gamma distributions are also described
in \cite{Knusel:1996:CNG}. 
The package {\bf ncg} includes functions for the computation of the noncentral gamma cumulative
distribution function (function {\bf pgammanc}), the computation of the noncentrality 
parameter (function {\bf deltagammanc}) and the
quantiles of the noncentral gamma distribution function (function {\bf qgammanc}). In our notation, this 
corresponds to the computation of the function $P_{\mu}(x,y)$, the inversion
of $P_{\mu}(x,y)$ with respect to $x$ and the the inversion
of $P_{\mu}(x,y)$ with respect to $y$, respectively. Therefore, there is a first difference
between our routines and those included in the package {\bf ncg}: our algorithms for the
direct computation allow to obtain both $P_{\mu}(x,y)$ and its complementary function, the function $Q_{\mu}(x,y)$. 
Also, the inversion algorithms compute for the minimum of $P_{\mu}(x,y)$ and $Q_{\mu}(x,y)$, which is very convenient
for computations in the queues of the distributions. 

In the computation of $P_{\mu}(x,y)$, we have checked that the function {\bf pgammanc} has some problems
 when computing small values of the cumulative
distribution function. This is illustrated in Table 3, where computations by {\bf pgammanc}, our routine {\bf cdfgamNC}
and the Mathematica function {\bf MarcumQ} for some values of $\mu$, $x$ and $y$ are shown.

 \begin{table}
$$
\begin{array}{c|c|c|}
\mu,\,x,\,y &	{\small \bf pgammanc}(y,\mu,2x) &   \\
\hline
                      &                                  & \mbox{GST: }  {\bf 1.215915354045}...{\bf 10^{-23}}   \\
   5,\,150,\,30       &   {\bf 1}.132428676150...{\bf 10^{-23}} & \\
              &                                       &   \mbox{MATH: } {\bf 1.215915354045}..{\bf 10^{-23}} \\
\hline
             &                           & \mbox{GST: }     {\bf 3.287840255874}...{\bf 10^{-30}} \\
1,\,75,\,0.5 & 1.215915354045...10^{-23} &    \\

  & &  \mbox{MATH: }  {\bf 3.287840255874}... {\bf 10^{-30}}     \\
\hline
   &   &  \mbox{GST: } {\bf 1.557081489535}...{\bf 10^{-35}} \\
2,\,100,\,2 &  1.976996751829...10^{-38} &  \\
   &  &  \mbox{MATH: }	{\bf 1.557081489535}...{\bf 10^{-35}} \\  
\hline 
   &   &   \mbox{GST: } {\bf 5.152185145235}...{\bf 10^{-48}} \\
10,\,100,\,1 &  5.486504449527...10^{-57} &	 \\
             &  &\mbox{MATH: }	{\bf 5.152185145235}... {\bf 10^{-48}}    \\
\end{array}
$$
{\footnotesize {\bf Table 3}.  Test of the routine {\bf cdfgamNC}: comparison of the values
obtained for $P_{\mu}(x,y)$ using the R function {\bf pgammanc}, our routine {\bf cdfgamNC} (values GST)
and the Mathematica function  $ N[{\bf MarcumQ}[\mu,\sqrt{2x},0,\sqrt{2y}]]$ (values MATH) for some values of $\mu$, $x$ and $y$.  }
\label{table3}
\end{table} 

Regarding the comparison of the inversion routine with respect to $x$ (the noncentrality
parameter), the computations using the 
function {\bf deltagammanc} included in the package {\bf ncg} seem to be very slow for
certain parameter regions (small $\mu$, large $y$, small $p$). For example, for computing $x$ in 
$P_{1.9}(x,288)=0.00001$ ($2x$ when calling the function {\bf deltagammanc})  
was about 40 $s$ when using {\bf deltagammanc} and $10^{-5}$ s when using {\bf invcdfgamNC}.

Another test for the inversion of the noncentral gamma distribution function,
is shown in Table 4. In this table, the relative error for the inversion
problem with respect to $x$ and $y$  of the noncentral gamma distribution function 
given in Eq.~(\ref{eq:int05}) for different
values of $(q,\,x)$ and $(q,\,y)$ ($\mu$ fixed to $0.5$) is shown. In a second test, $10^{7}$ 
random points have been generated in the domain of input parameters $\mu \in [0.5,\,10000]$,
$y \in [0,\,10000]$ ($x \in [0,\,10000]$) and $q \in [0,\,1]$ for testing the inversion
with respect to $x$ ($y$). 
Fixing the accuracy parameter of the secant method to $10^{-12}$, 
the accuracy of the computed  
values obtained in the inversion was always better than $10^{-11}$.

 \begin{table}
\label{table4}
$$
\begin{array}{cccc|ccc}
  &  & \mbox{Inv. wrt } x &   & & \mbox{Inv. wrt } y &  \\
  \hline
q & y=10 & y=100 & y=1000 & x=10 & x=100 & x=1000 \\
\hline
0.001 & 4\,10^{-13} & 4\,10^{-16}  & 1.5\,10^{-15} & 6.5\,10^{-16}   & 1.5\,10^{-15}  & 1.5\,10^{-15} \\
0.1 & 2\,10^{-15} & 8\,10^{-14}  &  4\,10^{-14}  & 2.8\,10^{-16}  & 8.1\,10^{-14} & 2.5\,10^{-13}\\
0.3 & 1\,10^{-13} & 0 & 3\,10^{-15} & 9.3\,10^{-16} & 1.5\,10^{-14}& 7.3\,10^{-14} \\
0.5 & 4 \,10^{-15} & 0  & 0  & 2.5\,10^{-14} & 0 & 2\,10^{-15} \\
0.7 & 5 \,10^{-15} & 8\,10^{-16}  &  4.9 \,10^{-15}  & 5\,10^{-15}  &   3.2\,10^{-14}  \\
0.999 & 2\,10^{-16} & 0 & 0  & 0  &  0  & 0 \\
\end{array}
$$
{\footnotesize {\bf Table 4}. Relative errors in the inversion of the noncentral gamma distribution function
with respect to (wrt) $x$ and with respect to $y$. The value of  $\mu$ has been fixed to $0.5$  }
\end{table}

The accuracy of the routines for the computation of the error function, the complementary error
function, the gamma function and its related functions have been tested by comparing
the values obtained against Maple (with a large number of digits). In all cases, a full agreement
in at least $13-14$ significant digits was found.   

Finally, Table 5 shows relative errors in the inversion process
of the complementary error function $y=\erfc (x)$ for different values of $y$.

 \begin{table}
$$
\begin{array}{c|c|c|}
y & \tilde{x} & \|\erfc(\tilde{x})/y-1\| \\
\hline
1.9 &  -1.1630871536766738 &  0\\
1.0 & -8.571414913863924\,10^{-17}   & 0 \\
10^{-2} &  1.163087153676674  &1.4\,10^{-16}  \\
10^{-3} & 2.3267537655135246  & 4.3\,10^{-16} \\
10^{-4} & 2.7510639057120607  & 2.7\,10^{-16} \\
10^{-5} & 3.123413274340875  & 1.2\,10^{-16}\\
10^{-6} & 3.4589107372795  & 8.4\,10^{-16} \\
10^{-7} &  3.766562581570838 & 6.6\,10^{-15} \\
10^{-8} &   4.052237243871389 & 2.1\,10^{-15}\\
10^{-9} & 4.320005384913445  & 8.3\,10^{-16}\\
10^{-10} & 4.572824967389486   & 4.8\,10^{-15}\\
10^{-11} &  4.812924067365823   & 4.8\,10^{-16} \\
10^{-12} & 5.042029745639059  & 0\\
\end{array}
$$
{\footnotesize {\bf Table 5}. Solutions $x(y)$ of the equation
$y=\erfc (x)$.  }
\label{table5}
\end{table}

\section{Test run description}

The Fortran 90 test program {\bf testgamCHI.f90} includes 
several tests for the computation
of function values and their comparison with the corresponding
pre-computed results. Also, the inversion routines are tested 
by checking that the composition of the function with its inverse is the identity.

\section{Appendix: The asymptotic inversion method}\label{sec:asinv}
The cumulative distribution functions considered in this paper can be written in one of the standard forms (see also \cite{Gil:2010:AIC})
\begin{equation}\label{eq:app01}
\begin{array}{@{}r@{\,}c@{\,}l@{}}
F_a(\eta)&=&\dsp{\sqrt{\frac{a}{2\pi}}\int_{-\infty}^\eta e^{-\frac12a\zeta^2}f(\zeta)\,d\zeta,}\\[8pt]
G_a(\eta)&=&\dsp{\sqrt{\frac{a}{2\pi}}\int_{\eta}^\infty e^{-\frac12a\zeta^2}f(\zeta)\,d\zeta,}
\end{array}
\end{equation}
where $a>0$, $\eta\in{{\RR}}$, and $f$ is analytic and real on ${{\RR}}$ with $f(0)=1$. The special case $f=1$ gives the normal distributions (see \eqref{eq:errorf03})
\begin{equation}\label{eq:app02}
\begin{array}{@{}r@{\,}c@{\,}l@{}}
P(\eta\sqrt{a})&=&\dsp{\sqrt{\frac{a}{2\pi}}\int_{-\infty}^\eta e^{-\frac12a\zeta^2}\,d\zeta=
\tfrac12{{\rm erfc}}\left(-\eta\sqrt{a/2}\right),}\\[8pt]
Q(\eta\sqrt{a})&=&\dsp{\sqrt{\frac{a}{2\pi}}\int_{\eta}^\infty e^{-\frac12a\zeta^2}\,d\zeta=
\tfrac12{{\rm erfc}}\left(+\eta\sqrt{a/2}\right).}
\end{array}
\end{equation}

We explain how the incomplete gamma function $P(a,x)$ defined in \eqref{eq:int02} 
can be written in this standard form. Let $\lambda=\frac{x}{a}$ and $t=a\tau$. Then
\begin{equation}\label{eq:app03}
\Gamma^*(a)P(a,x)=\sqrt{\frac{a}{2\pi}}\int_0^{\lambda} e^{-a(\tau-\ln \tau -1)}\,\frac{d\tau}{\tau},
\end{equation}
where $\Gamma^*(a)$ is defined in \eqref{eq:moc05}. The transformation 
\begin{equation}\label{eq:app04}
\tau-\ln \tau -1=\tfrac12\zeta^2,\quad {{\rm sign}}(\tau-1)={{\rm sign}}(\zeta),
\end{equation}
gives the standard form
\begin{equation}\label{eq:app05}
\Gamma^*(a)P(a,x)=\sqrt{\frac{a}{2\pi}}\int_{-\infty}^{\eta} e^{-\frac12a\zeta^2}f(\zeta)\,d\zeta,\quad f(\zeta)=\frac1{\tau}\frac{d\tau}{d\zeta},
\end{equation}
where $\eta$ is defined by
\begin{equation}\label{eq:app06}
\lambda-\ln\lambda-1=\tfrac12\eta^2,\quad {{\rm sign}}(\lambda-1)={{\rm sign}}(\eta).
\end{equation}

By using an integration by parts procedure in \eqref{eq:app01}  we can obtain an asymptotic representation of $F_a(\eta)$ which is valid for $a\to\infty$, uniformly for all $\eta\in{{\mathbb R}}$. We write $f(\eta)=[f(\eta)-f(0)]+f(0)$, where $f(0)=1$, and use \eqref{eq:app02}. Then we obtain by repeating integration by parts
\begin{equation}\label{eq:app07}
\begin{array}{@{}r@{\,}c@{\,}l@{}}
F_a(\eta)
&=&\dsp{\tfrac12{{\rm erfc}}(-\eta\sqrt{{a/2}})F_a(\infty)\ +\frac{e^{-\frac12a\eta^2}}{\sqrt{{2\pi 
a}}}S_a(\eta),}\\[8pt]
G_a(\eta)
&=&\dsp{\tfrac12{{\rm erfc}}(\eta\sqrt{{a/2}})G_a(-\infty)\ -\frac{e^{-\frac12a\eta^2}}{\sqrt{{2\pi 
a}}}S_a(\eta),}
\end{array}
\end{equation}
where $F_a(\infty)=G_a(-\infty)$ and
\begin{equation}\label{eq:app08}
F_a(\infty)\sim \sum_{n=0}^\infty \frac{A_n}{a^n}, \quad A_0=1,
\quad S_a(\eta)\sim \sum_{n=0}^\infty \frac{C_n(\eta)}{a^n},\quad a\to\infty,
\end{equation}
uniformly with respect to $\eta\in{{\mathbb R}}$.
The coefficients follow from the following recursive scheme. 
Let $f_0(\eta)=f(\eta)$. Then, for $n=0,1,2,\ldots$, define
\begin{equation}\label{eq:app09}
f_{n+1}(\eta)=\frac{d}{d\eta}\frac{f_{n}(\eta)-f_{n}(0)}{\eta},
\end{equation}
and we have
\begin{equation}\label{eq:app10}
A_n=f_n(0),\quad C_n(\eta)=\frac{f_n(0)-f_n(\eta)}{\eta}.
\end{equation}

To describe the inversion problem, we assume that $p\in(0,1)$ and that $a$ is a large positive parameter. Then we are interested in the value $\eta$ that solves the equation
\begin{equation}\label{eq:app11}
F_a(\eta)=F_a(\infty) \, p,
\end{equation}
where $F_a(\eta)$ has the form given in \eqref{eq:app07}.

First we define a number $\eta_0$ that solves the reduced equation 
\begin{equation}\label{eq:app12}
\tfrac12{{\rm erfc}}\left(-\eta_0\sqrt{{a/2}}\right) = p.
\end{equation}
Then for the desired value  $\eta$ we assume the expansion
\begin{equation}\label{eq:app13}
\eta\sim\eta_0+\frac{\eta_1}{a} +\frac{\eta_2}{a^2} +\frac{\eta_3}{a^3}+\ldots , \quad a\to\infty,
\end{equation}
and try to find the coefficients $\eta_1, \eta_2,\eta_3,\ldots$. To obtain the $\eta_j$ we can substitute the expansion for $\eta$ into the asymptotic expansion of $F_a(\eta)$\ and use formal power series manipulations. 

However, here we use the method as explained in our earlier mentioned publications. This method runs as follows. From \eqref{eq:app01},  \eqref{eq:app11}, and \eqref{eq:app12} we obtain
\begin{equation}\label{eq:app14}
\frac{dp}{d\eta_0}=\sqrt{\frac{a}{2\pi}} e^{-\frac12a\eta_0^2},\quad \frac{dp}{d\eta}=\sqrt{\frac{a}{2\pi}}\frac{f(\eta)}{F_a(\infty)}e^{-\frac12a\eta^2},\end{equation}
from which we obtain, upon dividing,
\begin{equation}\label{eq:app15}
f(\eta)\frac{d\eta}{d\eta_0}=F_a(\infty)e^{\frac12a\left(\eta^2-\eta_0^2\right)}.
\end{equation}
Substituting \eqref{eq:app13} and using $F_a(\infty)=1+\bigO(1/a)$, we obtain, after first-order perturbation analysis for large $a$,
\begin{equation}\label{eq:app16}
f(\eta_0)= e^{\eta_0\eta_1}\quad\Longrightarrow\quad \eta_1=\frac{1}{\eta_0}\ln f(\eta_0).
\end{equation}
Because $f$ is analytic at the origin with $f(0)=1$, $\eta_1$ is well defined as $\eta_0\to0$.

For higher-order terms $\eta_j, j\ge 2,$ we need in \eqref{eq:app15} more coefficients in the asymptotic expansion of $F_a(\infty)$ (see  \eqref{eq:app08}) and we have to expand
\begin{equation}\label{eq:app17}
f(\eta)=f(\eta_0)+(\eta-\eta_0)f^{\prime}(\eta_0)+\tfrac12(\eta-\eta_0)^2f^{\prime\prime}(\eta_0)+\ldots .
\end{equation}
Then, the next coefficients are given by
\begin{equation}\label{eq:app18}
\begin{array}{@{}r@{\;}c@{\;}l@{}}
\eta_2&=&\dsp{-\left(f\left(2A_1+\eta_1^2-2\eta_1^\prime\right)-2\eta_1f^\prime \right)/\left(2\eta_0f\right),}\\[8pt]
\eta_3&=&-\left(8f\eta_1\eta_2+4A_1f\eta_1^2+f\eta_1^4+4f\eta_1^2\eta_0\eta_2+4f\eta_0^2\eta_2^2-8f\eta_2^{\prime}+\right.\\[8pt]
&&\left.
8A_1f\eta_0\eta_2+8A_2f-8f^{\prime}\eta_1\eta_1^{\prime}-8f^{\prime}\eta_2-4f^{\prime\prime}\eta_1^2\right)/\left(8\eta_0f\right),
\end{array}
\end{equation}
where $f$, $f^\prime$ and $f^{\prime\prime}$ are evaluated at $\eta_0$.

For small values of $\eta_0$ (that is, when $p\sim\frac12$), we need expansions. We have
\begin{equation}\label{eq:app19}
\begin{array}{@{}r@{\;}c@{\;}l@{}}
\eta_1&=&\dsp{a_1+\tfrac12\left(2a_2-a_1^2\right)\eta_0+\tfrac13\left(3a_3-3a_1a_2+2a_1^3\right)\eta_0^2+\bigO\left(\eta_0^3\right),}\\[8pt]
\eta_2&=&\dsp{-\tfrac13a_1^3+2a_3+\tfrac18\left(-12a_2a_1^2+5a_1^4+24a_4\right)\eta_0+\bigO\left(\eta_0^2\right),}\\[8pt]
\eta_3&=&\dsp{\tfrac{1}{15}\left(4a_1^5-5a_2a_1^3-15a_3a_1^2+120a_5\right)+\bigO\left(\eta_0\right),}
\end{array}
\end{equation}
where $a_k$ are the coefficients in the expansion $\dsp{f(\zeta)=\sum_{k=0}^\infty a_k\zeta^k}$. The $a_k$ are related to the $A_k$ used in \eqref{eq:app18}:
\begin{equation}\label{eq:app20}
A_n=\left(\tfrac12\right)_n2^n a_{2n},\quad n=0,1,2,,\ldots.
\end{equation}

\section{Acknowledgements}

The authors acknowledge financial support from 
{\emph{Ministerio de Ciencia e Innovaci\'on}}, project MTM2012-11686. NMT thanks CWI, Amsterdam, for scientific support.



%% References

%\section*{References}

\bibliographystyle{elsarticle-num}
\bibliography{GammaCHI}



\end{document}